\documentstyle[seceq,epsf]{ptptex}

\title{General Relativistic Magnetohydrodynamic Simulations of\\
Black Hole Accretion Disks}

\author{John F. {\sc Hawley} and Jean-Pierre {\sc De Villiers}}

\inst{Department of Astronomy, University of Virginia, USA}

\recdate{ }

\abst{
Observations are providing increasingly detailed quantitative
information about the accretion flows that power such high energy
systems as X-ray binaries and active galactic nuclei.  Analytic models
of such systems must rely on assumptions such as regular flow geometry
and a simple, parameterized stress.  Global numerical simulations offer
a way to investigate the basic physical dynamics of accretion flows
without these assumptions.  For black hole accretion studies one
solves the equations of general relativistic magnetohydrodynamics. 
Magnetic fields are of fundamental importance to the structure and 
evolution of accretion disks because magnetic turbulence is the source of 
the anomalous stress that drives accretion.  We have
developed a three-dimensional general relativistic magnetohydrodynamic
simulation code to evolve time-dependent accretion systems
self-consistently.  Recent global simulations of black
hole accretion disks suggest that  the generic structure of the
accretion flow is usefully divided into five regimes:  the main
disk, the inner disk, the corona, the evacuated funnel, and the
funnel wall jet.  The properties of each of these regions are
summarized.
}

\begin{document}

\maketitle

\section{Introduction}

Disk accretion is one of the fundamental processes in astrophysics,
arising naturally from the combination of gravity and angular
momentum.  Accretion disks are expected to be found in a variety of
locations including semi-detached binaries, forming solar systems, and
in the nuclei of active galaxies.  Accretion into black holes has the
potential to produce the most energetic phenomena and, because black
holes have no upper limit in mass, black hole accretion is found in
active galactic nuclei  as well as in X-ray binary systems.  Although
the basic theory of black hole accretion was developed about thirty
years ago in a number of influential papers\cite{rf:LBP,rf:NT,rf:SS},
the full problem is complex, and we still lack a 
comprehensive understanding of accretion.   
The only solutions that can be obtained
analytically are highly simplified, relying upon time-stationarity and
spatial symmetry.  Increasingly detailed observations of black hole
systems are taxing the limits of these simple models.  In
particular, there is ample evidence that disks are anything but
time-stationary.  X-ray binary systems have several spectroscopically
distinct states, such as the low-hard and high-soft states, and can
rapidly transition between them.  X-ray timing experiments find
luminosity fluctuations on a wide range of timescales.  Some systems
have a preference for particular fluctuation frequencies called
quasi-period-oscillations (QPOs).  High frequency QPOs are comparable
to the frequency of the marginally stable circular orbit around a black
hole.  Quasars often show rapid variability, with relatively large
swings in their substantial total luminosities.  Black hole accretion
need not be highly luminous, however, and the black hole at the
Galactic center represents another extreme where, despite the gas rich
environment in which the hole is located, the luminosity is too low to
explain through a conventional accretion disk model.  Modeling is made
more difficult by the need to explain the rapid outbursts and flaring
that are also observed.

The goal of accretion disk modeling is to predict the appearance and
observational properties that would result for a certain black hole
mass and spin, given a value for the accretion rate.   For years
this goal was stymied by our lack of knowledge of the angular
momentum transport mechanism within disks, and theory had to make do with
adjustable parameters such as the Shakura-Sunyaev viscosity parameter
$\alpha$.\cite{rf:SS}  However, in the last decade, progress has been
both decisive and rapid.  The elucidation of the accretion disk
magnetorotational instability (MRI)\cite{rf:BH,rf:BH98} has led to an
emerging consensus that magnetic fields are (literally) the driving
force behind high energy accretion processes, whether the fields are
weak or strong.  Thus, accretion flows are {\em magnetohydrodynamical.}
The MRI allows magnetic field to tap into the available orbital and
gravitational energy by transporting angular momentum outward.  As a
consequence magnetic fields in the disk are amplified, and the
accretion flow is {\em turbulent}.  Some of the energy in this
turbulence will ultimately be dissipated at microscopic scales, where
it is converted into heat that can then be radiated.  Some of the
energy may also go into outflows and jets.

Obviously a turbulent, magnetized accretion disk is difficult to study
analytically.  It certainly does not become easier to model such a
disk if it orbits a rotating black hole.  To solve for the
time-dependent dynamics of magnetized plasmas in the relativistic
potential of Kerr black holes, one must turn to numerical simulations.
Interestingly, the first numerical simulations of accretion into a Kerr
black hole actually predate the above-cited papers on the classical analytic 
theory.  A general relativistic (GR) hydrodynamic code was developed by
Wilson in 1972.\cite{rf:Wil72}  A few years later he followed up with
the first GR-MHD simulations.\cite{rf:Wil}  Since those pioneering
days, there have been revolutionary advances in computer hardware
capabilities, and somewhat more incremental, albeit steady, advances in
finite differencing algorithms.  There are now several groups are
actively pursuing GR-MHD.  In this paper we will summarize the major
aspects of our specific effort, along with some results to date.

\section{Equations of GR-MHD}

The equations that we are solving are those of ideal GR-MHD consisting
of the law of baryon conservation, ${\nabla}_{\mu}\,(\rho\,U^{\mu}) =
0$, where $\rho$ is the baryon density, $U^{\mu}$ is the
four-velocity, and ${\nabla}_{\mu}$ is the covariant derivative, the
conservation of stress-energy, ${\nabla}_{\mu}{T}^{\mu\,\nu} = 0 $,
where ${T}^{\mu\,\nu}$ is the energy-momentum tensor for the fluid
(including the electromagnetic portion of the stress-energy), and the
induction equation, ${\nabla}_{\mu}{}^*{F}^{\mu\,\nu}= 0$, where
${}^*F^{\mu \nu}$ is the dual of the electromagnetic field strength
tensor.  Of course, these equations are just the starting point; the
difficulty one faces when developing a numerical solver is to select an
optimal formulation of the equations and a stable and robust algorithm
for evolving them.  The details of our code are given in De Villiers
and Hawley.\cite{rf:DH}  Here we provide a brief summary.

We are carrying out three-dimensional, operator-split, time-explicit
finite-differ\-encing on variables located at staggered points on a
nonuniform mesh.  We assume a stationary Kerr metric and work in 
Boyer-Lindquist coordinates.  The radial grid runs
from just outside the horizon (e.g., 2.05 $M$ for a Schwarzschild
hole) to an outer boundary at large radius (120 $M$), 
and is graded to concentrate zones in the inner regions.
We use first-order differencing in time.  We use a
non-conservative form of the equations and work directly with
particular forms of the 
baryon density, the internal energy, and the three spatial components of
the four momentum.  In particular,
the equation of momentum conservation is
\begin{eqnarray}\label{momcons}
\partial_t\left(S_j-\alpha\,b_j\,b^t\right)+
  {1 \over \sqrt{\gamma}}\,
  \partial_i\,\sqrt{\gamma}\,\left(S_j\,V^i-\alpha\,b_j\,b^i\right)+
  \nonumber\\
  {1 \over 2}\,\left({S_\epsilon\,S_\mu \over S^t}- 
  \alpha\,b_\mu\,b_\epsilon\right)\,
  \partial_j\,g^{\mu\,\epsilon}
 + \alpha\,\partial_j\left(P+{{\|b\|}^2 \over 2}\right) = 0 .
\end{eqnarray}
This is written in terms of the auxilliary four-momentum, $S_\mu =
(\rho\,h\ + {\|b\|}^2)\,W\,U_\mu$, where $h=1 + \epsilon + P/\rho$ is
the relativistic enthalpy, $W$ is the relativistic gamma factor
($=\alpha U^t$ where $\alpha$ is the lapse function), $b^\mu$ is the
magnetic field four-vector in the rest-frame of the fluid, and
${\|b\|}^2=g^{\mu\,\nu}\,b_\mu\,b_\nu$.  The momentum is subject to
the  normalization condition $g^{\mu \nu}\,S_\mu\,S_\nu =
-{(\rho\,h+{\|b\|}^2)}^2\,W^2$, which is algebraically equivalent to
the more familiar four-velocity normalization $U^\mu\,U_\mu=-1$.

Note that there are alternate ways to write the equation of momentum
conservation that, while analytically equivalent, correspond to
significant differences in numerical implementation.  One is to
difference directly the total stress energy $T^{\mu\nu}$ and solve
algebraically for the primitive variables.  This approach is taken by
Koide, Shibata, \& Kudoh\cite{rf:KSK}, Komissarov\cite{rf:Kom}, and
Gammie, McKinney, and T\'oth.\cite{rf:GMT}  There are advantages to
working with a conservative form of the equations, but there are also
disadvantages due to increased complexity and vulnerability to
numerical instability.  And while one can difference the equations
to ensure conservation of total momentum and energy, one is not
guaranteed a proper distribution of momentum among individual
components, nor energy among its many forms, e.g., kinetic, thermal,
and gravitational.  The advantages and disadvantages necessarily
inherent in any algorithm show clearly why the development of a
number of distinct codes is important.

One of the difficulties of the MHD equations is evolving the magnetic
field terms while maintaining the algebraic constraint $\nabla \cdot B
= 0$.  Some of the issues related to this constraint problem have been
reviewed by T\'oth\cite{rf:toth} We follow the \textit{constrained
transport} (CT) method of Evans and Hawley\cite{rf:EH}, who pointed out
that the constraint can be preserved in GR using a staggered mesh and
solving the induction equation in the form
\begin{equation}
F_{\alpha \beta , \gamma} +
F_{\beta  \gamma, \alpha} +
F_{\gamma \alpha, \beta } = 0.
\end{equation}
The CT magnetic field variables are defined
in terms of the electromagnetic field strength tensor  
$F_{\alpha\beta}$ as
\begin{equation}
{\cal{B}}^r      = F_{\phi \theta} \, , \,
{\cal{B}}^\theta = F_{r \phi} \, , \,
{\cal{B}}^\phi   = F_{\theta r} .
\end{equation}
The induction equation is completely determined by these three
variables when one assumes the flux-freezing condition,
$F^{\mu \nu}\,U_\nu = 0$.
The induction equation then reads
\begin{eqnarray}\label{ct}
\partial_j \left({\cal{B}}^j\right) & = 0 & \quad (\nu=0) ,\\
\label{ct.3b}
\partial_t \left({\cal{B}}^i\right) -
\partial_j \left(V^i\,{\cal{B}}^j-{\cal{B}}^i\,V^j\right)& = 0 &
\quad (\nu=i),
\end{eqnarray}
where $V^\mu = U^\mu/U^t$ is the transport velocity with $U^t =
W/\alpha$.  By staggering the placement of the ${\cal{B}}^i$ variables
on the grid one can easily enforce the divergence-free constraint
in the differencing scheme.\cite{rf:EH}

\section{Results}

There are a number of reasons why global turbulent disk simulations are
daunting, with or without general relativity.  One such reason is the
enormous range in length and time scales that characterize accreting
systems.  A disk in a binary system orbiting around a 10 $M_\odot$
black hole extends over several orders of magnitude in radius, from
$\sim 10^5$--$10^{9}$~m.  The radial extent can be even greater for a
disk orbiting a supermassive black hole in a galactic nucleus.  The
disk is generally confined to a region near the equator, with a
vertical thickness $H < r$.  The plasma inside the disk is
turbulent, and the turbulent length scales range from $\sim H$ to the
microscopic dissipation length.  Of course we cannot expect to resolve
the full turbulent cascade, nor should we need to do so.  The transport
properties of the turbulence should be dominated by the largest
length scales.  Still, it is clear that the scale height $H$ needs to be
well resolved if even a portion of the turbulent cascade is to be
fairly represented.  Now couple that issue to the fact that to reach a
self-sustaining, quasi-stationary state the simulations must be three
dimensional.  It is a well-known property of turbulence that it is
inherently three dimensional.  Flows that are restricted to two
dimensions feature inverse cascades of power to large scale.  Further,
no self-sustaining magnetic dynamo is possible in two dimensions.

In an explicit GR simulation the timescale is limited to be less than
the shortest light crossing time across a zone.  However, the dynamic
timescale in the disk is the Keplerian orbital frequency, which is
$\Omega_{Kep} = (GM/r^3)^{1/2}$.  Over two decades in radius the
orbital period varies by a factor of 1000.  Inside the disk, sound
speed, $c_s$, and Alfv\'en speed are generally less than the orbital
velocity $r\Omega$; average turbulent velocities $\delta v$ should be
even smaller.  The net ``drift'' velocity which accounts for the net
accretion, $v_{acc}$, is the long-term average of the radial turbulent
velocity, and hence it must be much less than $\delta v$.  The
difficulty is apparent:  the interesting disk evolution time is set by
the accretion timescale, and $v_{acc} \ll c_s \ll v_\phi < c$.  To
observe any significant disk evolution a very large number of timesteps
is required.

Since its development, we have used our GR-MHD code for a number of
three-dimensional accretion simulations around both Schwarzschild and
Kerr black holes.  The initial conditions consist of a torus of gas
orbiting the black hole.  We are primarily interested in studying
accretion disk structures that form self-consistently, and this
``isolated torus'' initial condition has the advantage that it is
independent of the boundary conditions.  The choice of initial magnetic
field topology may also have significant implications.  We have focused
on loops of field that have zero net flux when integrated over the
computational domain.  One alternative would be to study the flows that
develop in the presence of initial large-scale net magnetic
fields.\cite{rf:K03}   Large-scale net poloidal fields may prove to be
essential for strong jet formation, but simulations that begin with 
zero net magnetic flux can investigate the circumstances under
which such large scale fields might develop naturally in the accretion
flow. We have examined both poloidal and toroidal loops as initial field 
configurations.  We have evolved flows into Schwarzschild holes, and 
extreme Kerr holes, rotating in both the
prograde and retrograde sense.  We have carried out simulations
at several resolutions, and have evolved the resulting accretion flows
for thousands of $M$ in time (time is conventionally given in units of
$GM/c^3$).  We have not included radiative cooling, an assumption that
greatly simplifies the calculation, and makes our results most
applicable to systems where the accretion efficiency is low.  These
simulations are presented in a recent series of
papers.\cite{rf:DH2,rf:DHK,rf:HKDH}

One of the difficulties with simulations is distinguishing between
transient effects that are a consequence of the artificial initial
configuration, and effects that are more generic.  Of
course any initial condition is likely to result in a period of
transient behavior, and the evolution of a weakly magnetized initial
torus is no exception.  However, this initial evolution follows a
course of clearly identifiable and quite understandable phases.  First,
the magnetic field grows rapidly due to the MRI.  If there is a
significant radial field present it can also grow due to orbital
shear.  As the MRI saturates it promotes an immediate redistribution of
angular momentum within the torus towards a Keplerian power law.  This
redistribution causes the inner edge of the disk to move inward and the
outer portions to move outward.  The total simulation time in these
models was 10 orbits at a radius $r=25 M$ (about $8000 M$ in time).
Since the orbital times at the innermost stable circular orbits are
all less than $100M$, the disk that forms at small radius can undergo
an extended quasi-stationary evolution over many dynamic times, driven
by sustained internal MHD turbulence.  The outer disk continues to act
as a reservoir of material for this sustained accretion.

We have classified this quasi-steady accretion flow into five distinct
regions:  the main body of the disk, a magnetized coronal envelope, the
inner disk region near the marginally stable orbit, the evacuated axial
funnel region, and a funnel-wall jet located between the corona and the
evacuated funnel.  The physical characteristics of each of these
regions are as follows:

\noindent\textbf{Main disk:} The main disk is a somewhat thickened
disk, with a nearly constant opening angle.  In our simulations $H
\approx 0.2 r$,  although generally the thickness would be determined
by competing heating and cooling processes.  The disk is MHD turbulent,
with highly tangled magnetic fields; gas pressure dominates over
magnetic pressure.  The turbulence transports angular momentum outward
with an average stress level that is approximately equal to half the
total magnetic pressure.  This corresponds to values of $0.01$--$0.1$
in terms of the $\alpha$ parameter.  However, this stress is highly
variable in both time and space.  The global simulations done to date,
both in GR and with Newtonian\cite{rf:H00} or pseudo-Newtonian
gravity\cite{rf:HK01,rf:HK02,rf:HB} seem to obtain this generic
result:  the disk has a Keplerian power-law angular momentum
distribution throughout.  The specific angular momentum is everywhere
slightly sub-Keplerian in value, in keeping with a nonzero radial
pressure gradient and the disk's vertical thickness.

\noindent\textbf{Inner disk:}  While the main disk structure is
expected to span most of the radial extent of the accretion flow, at
some point the disk must change from Keplerian orbital motion to
plunging inflow into the black hole.  We label the region where this
occurs as the inner disk.  The details of the disk's inner region are
of considerable importance\cite{rf:KH} since the gravitational
potential is very steep near the black hole, and small radial
variations can mean a large difference in the total efficiency of
accretion.  The inner disk is marked by a local pressure and density
maximum that lies outside the location of the marginally stable
circular orbit.  Inside this pressure maximum, the disk begins to
transition from turbulence-dominated to inflow.  Beyond the marginally
stable orbit matter spirals in toward the event horizon, stretching
field lines as it falls.  Nothing special in the fluid variables or
their behavior marks the presence of the marginally stable orbit.
Pressure and density are continuous, and the stress does not go to
zero: the specific angular momentum in the fluid continues to decline.
The amount of stress acting on the gas here will also determine the
angular momentum input into the black hole, which has long-term
implications for the spin rate of the hole.  Work to date suggests that
enough angular momentum can be extracted prior to accretion to set the
black hole's spin equilibrium value at 0.9, substantially below the
maximum.\cite{rf:GSM}  Most of the matter in the plunging inflow
accretes into the black hole, but some is ejected into outflows into
the coronal envelope and along the funnel wall.  The accretion rate
into the hole is highly variable.  Further, the accretion rate drops
with increasing black hole spin.  It is possible that magnetic torques
driven by the hole's rotation supply angular momentum to the disk,
retarding the rate of accretion.

\noindent\textbf{Corona:} This is a region of low density above the
surface of the disk, where gas and magnetic pressure are comparable.
The corona forms from gas driven upward by vertical pressure
gradients.  Magnetic buoyancy also plays a role;  field rises into the
corona and the resulting coronal magnetic field is mainly
toroidal.  In the absence of cooling, the temperature of the disk
increases rapidly inward, and essentially the disk is too hot to be in
vertical equilibrium everywhere.   Gas lofted upward in the inner
regions finds that it can move radially outward and does so. However,
the outflowing coronal gas remains bound, unlike the gas in the funnel-wall
jet which is unbound and escapes to infinity.

\noindent\textbf{Funnel:} If the accreting gas has any angular
momentum, an evacuated axial funnel is unavoidable.  Even if the funnel
were somehow filled with gas, it would have to be either falling in or
heading out; equilibrium is not possible.  In the simulations, the
funnel is quite strongly evacuated of gas, but not of magnetic field.
Early in the evolution, as the first gas falls through the plunging
region, radial field is ejected from the disk into the funnel where it
establishes a more or less split monopole topology.  The field is
nearly force free, although its absolute strength is not huge.  The
magnetic pressure is in pressure balance with the surrounding corona.
In the case of a rotating Kerr hole the field is twisted by the
dragging of inertial frames, and a toroidal field forms.  There is a
net outward flux of energy in the field powered by the spin of the
hole.

\noindent\textbf{Funnel wall jet:}  Between the evacuated funnel and
the surrounding corona lies a thin region featuring a significant
unbound mass outflow.  We have designated this outflow as the funnel
wall jet.  These jets are highly dynamic and appear to be driven in
episodic bursts.  The typical outflow velocity within the jet is
$\approx 0.4c$.  The density across the jet drops sharply as one moves
into the funnel, so the main unbound mass flux is confined to a bi-conical
surface at the boundary between the funnel and the corona, hence its
designation as the funnel wall jet.  The jet originates deep in the
flow inside the last stable orbit, above the disk at the boundary
between the corona and the plunging region.  The launching of the jet
appears to be driven both by steep pressure gradients and by Lorentz
forces.  Pressure seems to play the primary role in driving material
into the jet, as the jet's mass flux variations seem to be driven by
fluctuations in the thickness of the inner disk.  The mass flux and
velocities in the jet are larger for black holes with greater spin,
suggesting that at least some of the power comes from the hole itself
through magnetic interactions.  However, the funnel wall jet is present
in Schwarzschild simulations and even in pseudo-Newtonian
simulations,\cite{rf:HB} so black hole spin is not essential.

\section{Conclusions}

Our results to date are, of course, only the first steps towards a more
comprehensive picture of the black hole accretion phenomena.  However,
we have identified several behaviors that appear to be quite general.
First, the MRI is present in the disk and produces the required angular
momentum transport to drive accretion.  The disks that result have a
Keplerian power-law angular momentum distribution, and the average
value of the specific angular momentum is everywhere slightly
sub-Keplerian, by an amount that depends on the thickness of the disk.
The disk is turbulent with an average stress that is about half the
magnetic pressure, which itself is well below the gas pressure in
value; toroidal fields dominate.  The disk is surrounded by a slowly
outflowing magnetized corona where the magnetic and thermal energies
are more nearly equal.  What one might call the inner edge of the disk
varies in time and spatial location, and the stress does not go to zero
at the marginally stable orbit.  In the absence of cooling, winds and
jets appear to be a natural by-product of the accretion flow.  The jet
is formed at the interface between the corona and the magnetically
dominated, gas-evacuated centrifugal funnel.  Jets from spinning holes
are more powerful, and the accretion rate into the hole drops with
increasing black hole spin.  There is evidence for energy and angular
momentum extraction from spinning holes. The maximally spinning holes
are actually being spun down; the specific angular momentum of the
material carried into the hole is not sufficient to maintain the spin
rate.

The astrophysical issues posed by these preliminary results should more
than occupy the time of the GR-MHD community in immediate future.  We
should also note that there remain a number of technical problems to
work through as well.  First, one of the challenges of these global
simulations is to extract and distill useful diagnostic information
that accurately characterizes the most important physical properties of
the system.  One computes, of course, a complete set of variables, at
all grid zones for all timesteps; choosing from this data what to
evaluate and to save is the issue.  Given the novelty of
three-dimensional GR-MHD accretion simulations, there is as yet no
definitive set of diagnostic quantities.  We anticipate that standards
will develop and evolve as our understanding of the simulation dynamics
of black hole accretion flows improves.

Second, all the current GR-MHD codes are quite new and need further
validation.  Part of this entails the development of a standard suite
of test problems, such as those used already.\cite{rf:DH,rf:GMT} It
will also be important to compare the detailed results from quite
different numerical algorithms.  So far we have made limited
comparisons with the axisymmetric results of Gammie et al.\cite{rf:GSM}
While these comparisons show excellent agreement between the two codes,
a deeper examination should provide valuable insights into the
strengths and weaknesses of the different numerical techniques.

\section*{Acknowledgements}
We wish to acknowledge the many contributions to this ongoing work
by our collaborators, Charles Gammie, Shigenobu Hirose,
and Julian Krolik.  This work was supported by NSF grants
AST-0070979 and PHY-0205155, and NASA grant NAG5-9266.

\end{document}